\documentclass[prd,aps,twocolumn,floats,floatfix,amsmath,nofootinbib]{revtex4-1}
\usepackage{graphicx,array,dcolumn}
\usepackage{calc,tabularx, epsfig,mathrsfs}
\usepackage{hyperref}

\usepackage{amsmath,verbatim,enumerate}
\usepackage{amssymb}
\usepackage{multirow}
\usepackage{xspace}
\usepackage{slashed}
\allowdisplaybreaks[1]
\newlength{\figurewidth}
\newcommand{\beq}{\begin{equation}}
\newcommand{\eeq}{\end{equation}}
\newcommand{\bea}{\begin{eqnarray}}
\newcommand{\eea}{\end{eqnarray}}
\newcommand{\ba}{\begin{array}}
\newcommand{\ea}{\end{array}}
\newcommand{\bg}{\bar{g}}
\newcommand{\mn}{{\mu\nu}}

\newcommand{\pt}{\partial}

\newcommand{\al}{\alpha}
\newcommand{\bt}{\beta}

\newcommand{\lam}{\lambda}
\newcommand{\Lam}{\Lambda}

\newcommand{\nb}{\nabla}

\newcommand{\D}{\Delta}

\newcommand{\om}{\omega}
\newcommand{\sg}{\sigma}
\newcommand{\kp}{\kappa}

\makeatother
\begin{document}
%
\title{Ultraviolet complete dark energy model}
\setlength{\figurewidth}{\columnwidth}
%
\author{Gaurav Narain$\,{}^a$}
\email{gaunarain@itp.ac.cn}
\author{Tianjun Li$\,{}^{a,b}$}
\email{tli@itp.ac.cn}
\affiliation{${}^a$ CAS Key Laboratory of Theoretical Physics, Institute of Theoretical Physics, 
Chinese Academy of Sciences, Beijing 100190, China\\
${}^b$ School of Physical Sciences, University of Chinese Academy of Sciences, 
No. 19A Yuquan Road, Beijing 100049, China}

%
%
\begin{abstract}
We consider a local phenomenological model to explain a
non-local gravity scenario which has been proposed to address dark energy issues. 
This non-local gravity action has been seen to fit the data as well 
as $\Lam$-CDM and therefore demands a more fundamental local treatment. 
The induced gravity model 
coupled with higher-derivative gravity is exploited for this proposal, as this 
perturbatively renormalizable model has a well-defined ultraviolet (UV) description 
where ghosts are evaded. We consider a generalised version of this model 
where we consider two coupled scalar fields and their non-minimal coupling with gravity. 
In this simple model, one of the scalar 
field acquires a Vacuum Expectation Value (VEV), thereby inducing a mass 
for one of the scalar fields and generating Newton's constant. The induced mass 
however is seen to be always above the running energy scale thereby leading to 
its decoupling. The residual theory after decoupling becomes a platform for driving 
the accelerated expansion under certain conditions. Integrating out the 
residual scalar generates a non-local gravity action. The leading term of 
which is the non-local gravity action used to fit the data of dark energy.    
\end{abstract}

\maketitle
%
%

\section{Introduction}
\label{intro}

Dark Energy (DE) has been a puzzling problem which is observed in the Universe 
in the era of large distance scale where the Universe is seen to be undergoing accelerated 
expansion \cite{Riess:1998cb,Perlmutter:1998np}. 
There has been several efforts in order to explain at a fundamental 
level this observed phenomenon of accelerated expansion at large cosmic distances: 
quintessence \cite{Wetterich:1987fm,Frieman:1995pm,Zlatev:1998tr,Brax:1999yv,Ferreira:1997au}, 
$\Lam$-CDM, K-essence \cite{Garriga:1999vw,ArmendarizPicon:1999rj,ArmendarizPicon:2000dh}. 
But so far $\Lam$-CDM seems to have the best fit 
with the data. However, a good fundamental explanation is currently still lacking. 

The idea of explaining accelerated expansion in the late time Universe with a 
cosmological-constant like term in the action goes back to the last century. 
However, the realisation that such a term could play a role in explaining the observed
accelerated expansion (as it give rise to negative pressure) is not very old. 
To explain the cause of this accelerated expansion which is obtained either by a constant energy-density 
term in the action or by some field whose energy density asymptotically approaches a
constant is known as dark energy. 

A good and simple way to get an accelerated expansion is by making use of a scalar field 
where its slow variation with respect to cosmic time give rise to negative pressure. 
This is the usual scenario in quintessence model 
\cite{Wetterich:1987fm,Frieman:1995pm,Zlatev:1998tr,Brax:1999yv,Ferreira:1997au}
and k-essence model \cite{Garriga:1999vw,ArmendarizPicon:1999rj,ArmendarizPicon:2000dh}. 
Of course, there are other procedures too which involves 
additional vector or bi-metric gravity. 
Recently an interesting proposal using non-locality has been 
suggested, as a way to get accelerated expansion 
\cite{Maggiore:2014sia,Cusin2016nzi,Maggiore2016gpx}. 
It is seen that for certain kind of non-locality the accelerated expansion 
achieved in late-time Universe fits the data as nicely as $\Lam$-CDM 
\cite{Dirian:2014bma,Dirian:2016puz}. 
However so far the explanation for the appearance of this 
non-locality does not exist, in the sense that it is not 
known whether it can arise from some fundamental local theories. 
Although it has been argued that similar kind of non-localities can arise 
in quantum theory where the energy dependence in the renormalisation 
group running of couplings has been generalised such that
$g(\mu) \to g(-\Box)$, where $\Box$ is the square of covariant
derivative while $g$ is some coupling of the theory 
\cite{Maggiore:2015rma,Maggiore:2016fbn}. Here the 
infrared behaviour of the running couplings is argued to lead to 
non-local modification of gravity. In this paper, we aim to address this issue 
of non-locality by proposing a local model where such a non-locality 
arises naturally when the fields decouple from the system. 
Such decoupling gives the non-local action which at late times results 
in accelerated expansion. 

The idea we exploit here is that in the higher-derivative induced gravity 
model, the quantum corrections give scalars an induced mass when scale-symmetry
is broken via Coleman-Weinberg procedure \cite{Narain2016}. 
However, the induced mass of scalar 
is seen to be always above the running energy scale 
resulting in a decoupling phenomenon 
\cite{Appelquist:1974tg, Gorbar:2002pw, Gorbar:2003yt,Gorbar:2003yp}.
In a model where there are more than one scalar, 
it is possible to achieve a phase of accelerated expansion at late times. 
In these cases if one of the scalar gets an induced mass after symmetry 
breaking and eventually decouples from the system, then the 
residual system can execute an era of accelerated expansion.
We show that integrating out the extra scalar from the residual 
action results in a non-local action whose leading term
matches with the non-local action considered by  
\cite{Cusin2016nzi, Dirian:2016puz, Maggiore2016gpx}. 

The outline of paper is as follows. In section \ref{nonlocal} we give a 
brief review of the non-local gravity. In section \ref{induced} we
give a brief outline of induced and higher-derivative gravity. 
In section \ref{model} we present the model which reduces 
to the non-local gravity model. Finally conclusion is 
presented in section \ref{conc}.

\section{Non-local Gravity}
\label{nonlocal}

Here in this section we will give a short review about the non-local gravity model which 
has been investigated in \cite{Cusin2016nzi, Dirian:2016puz, Maggiore2016gpx}. 
In this proposal the gravitational equation of motions are modified at long distances
by a non-local term. This was first suggested in \cite{ArkaniHamed:2002fu} where 
the General Relativity (GR) equations were phenomenologically modified to 
\beq
\label{eq:GRmodify}
\left(1 - \mu^2 \Box^{-1} \right) G_\mn = 8 \pi G T_\mn \, .
\eeq
Here $\mu^2$ dictates the characteristic length scale at which non-locality 
enters, $G_\mn$ is the Einstein tensor for the metric, $G$ is the 
gravitational coupling constant, and $T_\mn$ is the energy-momentum 
tensor. However, an immediate problem that arises here is that the 
energy-momentum tensor is not automatically conserved as the covariant derivative 
$\nb_\mu$ does not commute with $\Box^{-1}$ on a curved space-time. 
It was soon realised that if one demands that energy-momentum tensor $T_\mn$ 
should be covariantly conserved then one has to modify the equations 
of motions to
\beq
\label{eq:GRmodifyT}
G_\mn - \mu^2 \left(\Box^{-1} G_\mn \right)^T  = 8 \pi G T_\mn \, . 
\eeq
Here $T$ in the superscript denotes the transverse part of the tensor 
in brackets \cite{Jaccard:2013gla}. This modification implies that 
$T_\mn$ is conserved but it leads to unstable cosmological 
evolution \cite{Maggiore:2013mea,Foffa:2013vma}.
The first successful non-local model, free of instabilities 
was given by \cite{Maggiore:2013mea}.
\beq
\label{eq:GRmodifyTT}
G_\mn - \frac{\mu^2}{3} \left(g_\mn \Box^{-1} R \right)^T  = 8 \pi G T_\mn \, ,
\eeq
where the factor of $1/3$ was introduced to have a convenient normalisation 
for the mass parameter. In this model, there is no Veltman-Zakharov 
discontinuity and it smoothly makes a transition to GR when 
$\mu^2 \to 0$. Also, at cosmological scales, its evolution is 
stable during radiation and matter dominated era. At later 
times the non-local term behaves as dark energy giving rise
to an accelerated expansion \cite{Maggiore:2013mea,Foffa:2013vma}.
Moreover, its cosmological perturbations are well behaved both 
in scalar \cite{Dirian:2014ara} and tensor sector 
\cite{Dirian:2016puz, Cusin:2015rex}. Further investigations of this 
model reveals that it is consistent with CMB, supernova, Baryon 
Acoustic Oscillations (BAO) and structure formation data
\cite{Dirian:2014ara,Nesseris:2014mea,Barreira:2014kra}. 
A detailed comparison with $\Lam$-CDM shows that 
this non-local model fits the data at a level which is statistically 
indistinguishable from $\Lam$-CDM \cite{Dirian:2014bma,Dirian:2016puz}.
This was named $RT$-model 
where $R$ stands for Ricci Scalar while $T$ refers to 
transverse part. 

Currently an action for (\ref{eq:GRmodifyTT}) doesn't exist. 
However, this model is closely related to following non-local action,
\beq
\label{eq:nonLocalGR}
S_{\rm NL} = \frac{M_P^2}{2} \int \, {\rm d}^4x \sqrt{-g} 
\biggl[
R - \frac{\mu^2}{6} R \Box^{-2} R
\biggr] \, ,
\eeq
where $M_P$ is the reduced Planck mass. This action generates 
equation-of-motions (EOM) which on linearising 
around flat space-time matches the one obtained after linearising 
Eq. (\ref{eq:GRmodifyTT}). However at the non-linear level 
the two models are different. This model works very well 
at background level \cite{Maggiore:2014sia} and matches nicely with data
\cite{Dirian:2014ara,Dirian:2016puz} 
(although it does not fit as well as $\Lam$-CDM). 
This model is known as the $RR$-model
(see \cite{Nersisyan:2016hjh} for its dynamical system analysis). 

However, there are some arbitrariness in the sense of describing 
the non-local model given by Eq. (\ref{eq:GRmodifyTT})
at the level of the action. This gives one freedom to consider 
various kinds of non-local actions. But the restrictions
coming from stable cosmic evolution limit the number 
of allowed terms in the non-local modified action. It was 
shown that the following non-linear extension 
\beq
\label{eq:nonLocalconfGR}
S_{\rm NL} = \frac{M_P^2}{2} \int \, {\rm d}^4x \sqrt{-g} 
\biggl[
R - \frac{\mu^2}{6}  R \left(-\Box + \xi R \right)^{-2} R
\biggr] \, ,
\eeq
(where $\xi$ is a dimensionless parameter) does 
a very good job in matching the data. Such a non-linear 
extensions are inspired from the realisation that the $RT$-model
is a non-linear extension of $RR$-model \cite{Cusin2016nzi}. 
This model has been studied extensively and can 
fit the DE data nicely (statistically as well as 
$\Lam$-CDM). 

\section{Induced Gravity Model}
\label{induced}

Here we present a small review of the induced gravity 
model where we consider a non-minimally coupled scalar field 
with higher-derivative gravity action. The higher-derivative gravity 
action we consider is of the fourth order. The quantum theory of this 
is known to be renormalizable to all loops \cite{Stelle1977c, Stelle77}, 
and was recently shown to be unitary \cite{NarainA1, NarainA2} 
(see also references therein). This then offers a sufficiently  
simple quantum field theory of gravity which can be 
used to investigate physics at ultra-high energies. 

This scale-invariant model is like an induced gravity model \cite{Narain2016}, where a 
scalar acquires a VEV and in turn gives rise to the gravitational coupling
as well as generating masses for other fields. 
The scale-invariant system consists of only dimensionless couplings.
This makes the theory perturbatively renormalizable to all loops 
in four-dimensional space-time by power-counting \cite{Fradkin1981, Fradkin1982}
(for classical picture of these theories see \cite{Stelle1977c, Gaume2015}).
Scale-invariant gravitational systems coupled with matter 
have been investigated in the past. Some of the first studies were
done in \cite{Julve1978,Fradkin1981, Fradkin1982,Barth1983, Avramidi1985}, 
where the renormalisation group running of various couplings was computed 
and fixed point structure was analysed. Further investigation 
for more complicated systems were done in 
\cite{Odintsov1989,Buchbinder1989,Shapiro1989,Elizalde1994,
Elizalde1995_1,Elizalde1994_2,deBerredoPeixoto2003,deBerredoPeixoto2004}
(see also the book \cite{Buchbinder1992} for more details).

Recently the topic has gained some momenta and these models 
have been reinvestigated \cite{Strumia1,Einhorn2014,Jones1,Jones2,Jones2015}.
The purpose of these papers was to see if it is possible to generate 
a scale dynamically starting from a scale-invariant system. In 
\cite{Strumia1} the authors called their model `Agravity', where 
the Planck scale is dynamically generated from the VEV 
of a potential in the Einstein frame (not the Jordan frame).
They achieve a negligible cosmological constant, 
generate the Planck's scale, and addresses naturalness \cite{Strumia1, Salvio2016} and 
inflation \cite{Kannike2015}, but unitarity issues were not explored
\footnote{
In \cite{Salvio2015} a quantum mechanical treatment of 
$4$-derivative theories was suggested, which when suitably extended 
can tackle more complicated field theoretic systems. This can 
perhaps address issues of ghosts and unitarity in a more 
robust manner.}. 
In \cite{Narain2016} it was realised that the induced mass of the 
ghost is always above the energy scale and hence is innocuous. 
In \cite{Einhorn2014,Jones2015,Jones1,Jones2}
the authors studies the issue of dynamical generation of 
scale via dimensional transmutation in the presence of 
background curvature. This also induces Einstein-Hilbert 
gravity and generates Newton's constant, but the unitarity 
problem was not addressed. An interesting idea 
has been suggested in \cite{Holdom2015,Holdom2016}
by assuming an analogy with QCD, where the authors addressed 
the problem of ghosts and tachyons using the wisdom 
acquired from non-perturbative sector of QCD, as
it is argued that the gravitational theory enters a non-perturbative regime
below the Planck scale.

The idea of induced gravity goes long back. It was first proposed in
\cite{Zeldovich1967,Sakharov1967}, where the quantum matter 
fluctuations at high energy generate gravitational dynamics at 
low energy inducing the cosmological and Newton's gravitational constant. 
Another proposal suggested in \cite{Fujii1974, Chudnovsky1976,Zee1978} induces Einstein 
gravity spontaneously via symmetry breaking along 
the lines of the Higgs mechanism. Later in \cite{Zee1980,Adler19801, 
Adler19802, Adler19803, Adler1982} the idea of generation of Einstein 
gravity via dynamical symmetry breaking was considered, following 
the methodology of Coleman-Weinberg \cite{Coleman1973}. 
In \cite{Adler1982}, metric fluctuations were also incorporated 
in the generation of induced Newton's constant. Around the same 
time an induced gravity from Weyl-theory was studied 
\cite{Nepomechie1983, Zee1983, Buchbinder1986,Elizalde:1992vi, Odintsov:1991nd} as well.
Phase-transitions leading to the generation of 
Einstein-Hilbert gravity due to loop-effects from a conformal 
factor coupled with a scalar field were studied in 
\cite{Shapiro1994}. In \cite{Floreanini1993,Floreanini1994}
the renormalization group improved effective-potential 
of the dilaton leads to running of VEV thereby inducing mass scale (along with 
Einstein-Hilbert gravity). Furthermore, the authors make a proposal  
along lines of \cite{Salam1978,Julve1978} to tackle ghost and tachyons. 

The renormalizable and UV well defined scale-invariant action that one  
considers here is 
\bea
\label{eq:Act}
&&
S_{\rm GR} = \int {\rm d}^4x \sqrt{-g} \biggl[ \frac{1}{16\pi} \biggl\{
- \frac{1}{f^2}\left(R_\mn R^\mn - \frac{1}{3}R^2 \right) 
\notag \\
&&
+ \frac{\om}{6f^2} R^2 \biggr\}
+ \frac{1}{2} \pt_\mu \phi \pt^\mu \phi - \frac{\lam}{4} \phi^4 - \frac{\xi}{2} R \phi^2 
\biggr] \, ,
\eea
where the coupling parameters $f^2$, $\om$, $\lam$ and $\xi$ are all 
dimensionless, and the geometric quantities (curvature and covariant-derivative) 
depend on metric $g_\mn$. If one decomposes the metric 
$g_\mn = \bg_\mn + h_\mn$, where $\bg_\mn$ is the background metric 
and $h_\mn$ is the fluctuation, then one can compute the propagator of the 
fluctuation field $h_\mn$ and its various couplings. If the background 
metric is flat then the quantum metric fluctuations in momentum 
space (for the Landau gauge condition $\pt^\mu h_\mn=0$) is given by,
\beq
\label{eq:GR_prop}
D^{\mn \rho\sg} = (\D_G^{-1})^{\mn\rho\sg} 
=  (16 \pi) \frac{f^2}{q^4} \left(
- 2 P_2^{\mn \rho\sg}
+ \frac{1}{\omega} P_s^{\mn\rho\sg} \right) , 
\eeq
where $P_2^{\mn \rho\sg}$ and $P_s^{\mn\rho\sg}$ are spin 
projectors. They can be written in flat space-time in momentum space 
in a simple form given by
\bea
&&
\label{eq:spin2proj}
(P_2)_{\mu\nu}{}^{\alpha \beta}
 = \frac{1}{2} \left[ T_{\mu}{}^{\alpha} T_{\nu}{}^{\beta} + 
T_{\mu}{}^{\beta}T_{\nu}{}^{\alpha} \right] - \frac{1}{d-1} T_{\mu\nu}T^{\alpha\beta} \, ,
\\
&&
\label{eq:spinsproj}
(P_s)_{\mu\nu}{}^{\alpha \beta} 
= \frac{1}{d-1} T_{\mu\nu} \, T^{\alpha \beta} \, ,
\eea
where 
\begin{gather}
\label{eq:LTproj}
L_{\mu\nu} = \frac{q_{\mu} \, q_{\nu}}{q^2} \, \ , \hspace{10mm}
T_{\mu\nu} = \eta_{\mn} - \frac{q_{\mu} \, q_{\nu}}{q^2} \, . 
\end{gather}
These are basically the projector for projecting out various components
of a vector field. They satisfy $q^\mu T_\mn=0$ and $q^\mu L_\mn=q_\nu$.
Using them the projectors for the rank-2 tensor field 
can be constructed. The sign of couplings are taken in such a way so that 
the system doesn't generate any tachyons after the symmetry breaking
\cite{Narain2016}. This implies that $f^2>0$, $\om>0$, $\lam>0$ and 
$\xi>0$ (for the details on this choice of signs see \cite{Narain2016, Narain:2017tvp}). 
For these choice of signs the system remains stable and tachyons free.

Due to quantum corrections a VEV is generated in the effective potential of the 
scalar field, which then becomes a new vacuum. The original $\varphi^2=0$ 
vacuum becomes unstable under quantum corrections and the field migrates to the 
new vacuum which occurs at $\varphi^2=\kp^2$. It is given by \cite{Narain2016}
\beq
\label{eq:Veffmin}
\left. \frac{{\rm d}}{{\rm d}\varphi^2} Re(V_{\rm eff}) \right|_{\varphi^2 = \kp^2} = 0 \, .
\eeq
The generation of VEV consequently gives mass to scalar and generates 
an effective Newton's constant with the right sign, if the parameter $\xi$ was of opposite 
sign then the Newton's constant generated is of wrong sign \cite{Narain2016}. 
The generated mass and Newton's coupling can be expressed in terms of 
VEV $\kp^2$ and all the other couplings as
\beq
\label{eq:symbr_massG}
m_s^2 = \frac{3}{2} \lam \kp^2 \, , \hspace{5mm}
G^{-1} = 8 \pi \xi \kp^2 \, .
\eeq
The graviton propagator after the symmetry breaking is following \cite{Narain2016} 
\bea
\label{eq:GRprop_sb}
&&
D^{\mn,\al\bt} = 16 \pi G\cdot \Biggl[
\frac{(2P_2 - P_s)^{\mu\nu, \alpha\beta}}{q^2 + i \, \epsilon}
+ \frac{(P_s)^{\mu\nu, \alpha\beta}}{q^2 - M^2/\omega + i \epsilon}
\notag \\
&&
- \frac{2 \, (P_2)^{\mu\nu, \alpha\beta}}{q^2 - M^2+ i \epsilon}
\Biggr] \, ,
\eea
where now $G$ is the induced Newton's constant and is defined 
using Eq. (\ref{eq:symbr_massG}). The masses $M^2$ and $M^2/\om$ 
are given by
\beq
\label{eq:M2exp}
M^2 = 8 \pi f^2 \cdot \xi \kp^2 \, , 
\hspace{5mm}
\frac{M^2}{\om} = 8 \pi \frac{f^2}{\om} \xi \kp^2 \, .
\eeq
From here we immediately note that if $\om<0$ then there will be tachyons 
in the theory signalling an instability. The only way this catastrophe is avoided 
is when the parameter $\om>0$ \cite{NarainA1,NarainA2,Narain2016,Narain:2017tvp}.

From the propagator in the broken phase (\ref{eq:GRprop_sb}) we realise 
that the last term has a wrong sign leading to trouble with 
unitarity. This is the consequence of having higher-derivatives term in the action, 
resulting in ghosts. 
The generation of mass for the various modes allows one to 
investigate whether the induced mass for this ghost and its subsequent running 
can be such that it is always above the running energy scale. 
If the running of the parameters in the theory is such that the induced mass 
of the ghost remains always above the running energy scale, then the ghost mode 
never gets excited during the RG flow of the couplings In the case of pure 
higher-derivative gravity without matter \cite{NarainA1,NarainA2}, it was 
indeed noticed from the RG running of parameters that there exists a large 
domain of coupling parameters where the ghost mass is always 
above the RG energy scale, and is innocuous. This phenomena was also 
witnessed in the case of induced gravity model investigated in 
\cite{Narain2016}. The range of energy where the ghost remains 
physically unrealisable is quite large, extending from from very high 
in the ultraviolet (at or more than Planck's scale) to very deep in 
infrared (almost cosmological scales). This range of energy is set by the 
parameter $\om$ within which it remains positive and changes sign outside 
this range. For the system to be stable and tachyons free it is required that 
the parameter $\om$ should remain positive \cite{NarainA1,NarainA2}. 
The signs and stability of such higher-derivative systems have been discussed 
in more details in \cite{Narain:2017tvp}. 

In the case of induced gravity \cite{Narain2016}, although the induced mass 
of the ghost is such that it is always above the energy scale, the occurrence of
this phenomena leads to an interesting behaviour for the scalar. The induced scalar mass is 
seen to be always above the energy scale leading to its decoupling
in the same manner as as for the higher-derivative ghost.
It turns out that if one avoids ghosts to make the theory unitary then 
under appropriate choice of parameters, it is seen that 
the scalar gets also decoupled from the system. This decoupling of 
the scalar will have natural consequences in cosmology. In the following,
we consider a generalisation of this induced gravity action by incorporating 
an additional scalar field. One of the scalar gets a VEV and generates Newton's 
constant and masses, and gets decoupled. The left-over scalar in the 
residual action plays an interesting dynamics.

\section{Model}
\label{model}

In this section we present the idea of how the non-local action 
given in Eq. (\ref{eq:nonLocalconfGR}) can possibly arise from a 
local model via decoupling. 
This kind of non-local action can be safely embedded in the induced gravity scenario, 
which has an elegant UV completion \cite{Narain2016}. 
The simple idea is that in the higher-derivative induced gravity action 
the scalar field gets completely decoupled from the system 
under certain circumstance when the ghosts are avoided. 
This decoupling can then be exploited to construct models 
which can give rise to non-localities in far infrared. 
As this model has a sensible UV completion, 
it becomes a good unified picture explaining some of the 
UV and IR physics together.  

We consider the following two-scalar field model coupled non-minimally 
to gravity,
\bea
\label{eq:coupleAct}
&&
S = \int \, {\rm d}^4x \sqrt{-g} \biggl[
\frac{1}{2} \Phi^T \mathbf{\xi} \Phi R
+ \frac{1}{2} \Phi^T (-\Box) \Phi 
- V(\Phi^T \Phi) \biggr] 
\notag \\
&&
+\int \, {\rm d}^4x \frac{\sqrt{-g}}{16\pi} \biggl[
- \frac{1}{f^2}\left(R_\mn R^\mn - \frac{1}{3}R^2 \right) 
+ \frac{\om}{6f^2} R^2 \biggr] \, ,
\eea
where $\Phi = \{\phi, \chi \}$ is a two real scalar field doublet. This model 
is an extension of the induced gravity model stated in Eq. (\ref{eq:Act}), 
where the gravitational couplings are taken to have the same sign as before 
in order to ensure stability of system by no generating tachyons. The parameter 
$\mathbb{\xi}$ is now a matrix whose entries are given by,
\beq
\label{eq:xiM}
\mathbb{\xi} = \left(
\begin{array}{c c}
\xi_1 & \xi_{12} \\
\xi_{21} & \xi_2 
\end{array}
\right) \, ,
\eeq
where the entries of this matrix are all dimensionless. 
This is a simple two-scalar field model where the 
two scalar fields are not only coupled with each other but also have 
non-minimal coupling with gravity. It is a scale-invariant action 
which is renormalizable to all loops \cite{Stelle77,Fradkin1981,Fradkin1982}
\footnote{higher-derivatives gravity terms ensures renormalizability of theory
which remains unaffected by inclusion of renormalizable matter couplings}.
As this model is just a minor extension (by inclusion of an additional scalar) 
of the induced gravity model stated in 
Eq. (\ref{eq:Act}) where ghosts are evaded \cite{Narain2016}, so this model 
inherits the same virtues of the induced gravity model in which higher-derivative 
ghost are innocuous. 

The potential is taken to be of $\phi^4$ type as in four dimensions 
it is the only allowed term which obeys scale symmetry and 
renormalizability (similar kind of models were also considered in 
\cite{GarciaBellido:2011de,Karananas:2016kyt}). 
The form of potential is given by
\beq
\label{eq:Potscale}
V(\Phi^T \Phi) = \frac{1}{4} (\Phi^T \cdot \mathbb{\lam}\cdot \Phi)^2 \, 
\eeq
where the coupling $\mathbb{\lam}$ is a matrix 
\beq
\label{eq:lamcoup}
\mathbb{\lam} = \left(
\begin{array}{c c}
\lam_1 & \lam_{12} \\
\lam_{21} & \lam_2 
\end{array}
\right) \, 
\eeq
consisting of only dimensionless entries. 
In this model when the scalar $\phi$ acquires a VEV then it will generate 
mass terms for the various fields and will induce Newton's constant. Here 
we will not discuss the process through which VEV for $\phi$ gets 
generated but we assume that scale-symmetry 
breaking has already occurred resulting in a generation of nonzero VEV
$\kp$ for the scalar $\phi$. The paper \cite{Narain2016} gives a detailed 
description of generation of VEV via a Coleman-Weinberg procedure, 
where the quantum fluctuations leads to symmetry breaking. 
The fluctuation of $\phi$ around the VEV 
are denoted by $\varphi$. Here in this paper we are interested in the aftermath of the
symmetry breaking mechanism 
to know the behaviour of this model at late times. 
The higher-derivatives gravity terms can be ignored as in the deep infrared  
they do not play a role and gets heavily suppressed \cite{Narain2016}. 
This action can be written in an alternative form after opening up the matrix as follows 
\bea
\label{eq:Sact_open}
S &=& \frac{1}{2} \int \, {\rm d}^4x \sqrt{-g} \biggl[
\phi (-\Box + \xi_1 R) \phi
+ \chi (-\Box + \xi_2 R) \chi
\notag \\
&&
+ \phi(-2\Box + \xi_{12} R + \xi_{21} R )\chi
\notag \\
&&
-\frac{1}{2} \left\{\lam_1 \phi^2 + (\lam_{12} + \lam_{21}) \phi \chi + \lam_2 \chi^2 \right\}^2
\biggr] \, .
\eea
If $\phi$-field acquires a VEV then it generates 
Newton's constant giving rise to the usual gravitational dynamics. 
However the fluctuation field $\varphi$ around the VEV of $\phi$
couples with the other scalar field $\chi$. 
The action after symmetry breaking is given by,
\bea
\label{eq:symbreakAct}
&& 
S = \frac{1}{2} \int \, {\rm d}^4x \sqrt{-g} \biggl[
\kp^2 \xi_1 R + 2 \kp \xi_1 \varphi R + \varphi (-\Box + \xi_1 R) \varphi 
\notag \\
&&
+ \kp (\xi_{12} + \xi_{21})R \chi
+ \varphi(-2\Box + \xi_{12} R + \xi_{21} R )\chi
\notag \\
&&
+ \chi (-\Box + \xi_2 R) \chi
-\frac{1}{2} \bigl\{
\lam_1 \kp^2 + 2 \lam_1 \kp \varphi + (\lam_{12} + \lam_{21}) \kp \chi
\notag \\
&&
+ \lam_1 \varphi^2 + (\lam_{12} + \lam_{21}) \varphi \chi 
+ \lam_2 \chi^2 
\bigr\}^2
\biggr] \, .
\eea
The potential piece written in the second line when expanded will generate mass terms 
for the fields beside generating interactions. The mass term 
can be written in an elegant form in matrix notation as follows
\bea
\label{eq:Massmat}
&&
- \frac{1}{2} \left(
\begin{array}{c c}
\varphi & \chi 
\end{array}
\right) \cdot \mathbb{M} \cdot 
\left(
\begin{array}{c}
\varphi \\
\chi
\end{array}
\right) 
=
- \frac{1}{4} \kp^2 
\left(
\begin{array}{c c}
\varphi & \chi 
\end{array}
\right) 
\cdot
\notag \\
&&
\left(
\begin{array}{c c}
6\lam_1^2 &  \lam_1 (\lam_{12} + \lam_{21}) \\
\lam_1 (\lam_{12} + \lam_{21}) & (\lam_{12} + \lam_{21})^2 + 2 \lam_1 \lam_2 
\end{array}
\right) 
\cdot 
\left(
\begin{array}{c}
\varphi \\
\chi
\end{array}
\right) \, .
\eea
The interactions are given by
\bea
\label{eq:intact}
&&
I = -\frac{1}{4} \lam_1^2 \kp^4 - \frac{\lam_1 \kp^3}{2} \left\{2\lam_1 \varphi
+ (\lam_{12} + \lam_{21})\chi \right\}
\notag \\
&&
- \frac{\kp}{2} \bigl\{
2 \lam_1^2 \varphi^3 + 3 \lam_1 (\lam_{12} + \lam_{21}) \chi \varphi^2
\notag \\
&&
+ ((\lam_{12} + \lam_{21})^2 + 2 \lam_1 \lam_2) \varphi \chi^2
+ (\lam_{12} + \lam_{21}) \lam_2 \chi^3
\bigr\}
\notag \\
&&
- \frac{1}{4} \bigl \{
\lam_1^2 \varphi^4 + ((\lam_{12} + \lam_{21})^2 + 2 \lam_1 \lam_2) \varphi^2 \chi^2 
\notag \\
&&
+ 2 \lam_1 (\lam_{12} + \lam_{21}) \varphi^3 \chi 
+ 2 \lam_2 (\lam_{12} + \lam_{21}) \varphi \chi^3 
+ \lam_2^2 \chi^4 
\bigr\} \, .
\eea
Here the first term is a like a vacuum energy term, the linear in fields 
will give rise to tadpoles in the quantum theory which can be absorbed via 
field redefinitions, the cubic and quartic interactions will give rise to 
non-linear interactions. In special scenario where $(\lam_{12} + \lam_{21})=0$,
the mass matrix acquires a simple diagonal form while also simplifying the 
interaction pieces. We will consider this special case in the following. 
For this special case, the fields have the following masses (no mixing),
\beq
\label{eq:massSP}
m_1^2 = 3 \lam_1^2 \kp^2 \, ,
\hspace{10mm}
m_2^2 = \lam_1 \lam_2 \kp^2 \, .
\eeq
These are the induced masses for the fields $\varphi$ and $\chi$. 
The interaction piece for this special case is
\bea
\label{eq:intsimp}
&&
I = -\frac{1}{4} \lam_1^2 \kp^4 - \lam_1^2 \kp^3 \varphi
- \lam_1 \kp \varphi \bigl\{
\lam_1 \varphi^2 + \lam_2 \chi^2 
\bigr\}
\notag \\
&&
- \frac{1}{4} \bigl \{
\lam_1^2 \varphi^4 + 2 \lam_1 \lam_2 \varphi^2 \chi^2 
+ \lam_2^2 \chi^4 
\bigr\} \, .
\eea
One can then compute the equation of motion 
for the fields $\varphi$ (fluctuation) and $\chi$ respectively by varying the 
full residual action as follows
\bea
\label{eq:eqm}
&&
(-\Box + \xi_1 R - m_1^2) \varphi + (\xi_1 R - \lam_1 \kp^2) \kappa
\notag \\
&&
+ \frac{1}{2} (-2\Box + \xi_{12} R + \xi_{21} R)\chi 
-\lam_1 \kp(3 \lam_1 \varphi^2 + \lam_2 \chi^2)
\notag \\
&&
- \lam_1 \varphi (\lam_1 \varphi^2 + \lam_2 \chi^2) = 0 \, ,
\\
&&
(-\Box + \xi_2 R - m_2^2) \chi + \frac{1}{2} (\xi_{12} + \xi_{21} ) \kappa R
\notag \\
&&
+ \frac{1}{2} (-2\Box + \xi_{12} R + \xi_{21} R)\varphi 
- 2 \lam_1 \lam_2 \kp \varphi \chi 
\notag \\
&&
- \lam_2 \chi (\lam_1 \varphi^2 + \lam_2 \chi^2) = 0 \, ,
\eea
where each equation contains linear and non-linear interactions terms. 
These are coupled differential equations. 
If the interaction strength is small (which is the case in cosmological scenarios)
then one can solve the equations perturbatively. Moreover, in the scenario when 
the scalar $\varphi$ gets entirely decoupled from the system, the set of equations 
are greatly simplified. This is the approximation where mass $m_1^2$ is very large. 
In the induced gravity, higher-derivative model considered in 
Eq. (\ref{eq:Act}), such a phenomena naturally occurs when 
the scale-symmetry is broken. This breaking of scale 
symmetry not-only induces the gravitational coupling 
(Newton's constant) but also induces mass for the 
various fields. Under the renormalisation group running it is seen that the induced 
mass of the scalar is always above the running energy scale
(under certain conditions \cite{Narain2016}), resulting in 
a decoupling phenomena. This means that $m_1^2/E^2>1$
($E$ is the running energy). This condition means that 
the particle never goes on-shell. In infrared it is seen that 
this ratio $m_1^2/E^2 \gg 1$, which will imply that 
$m_1^2 \gg \Box$. This also means that in quantum theories 
when such a particle appears inside a loop then it does not
contribute to the imaginary part of the forward scattering 
amplitude (Cutkowsky cut), as it never goes on-shell. 
Here in this paper we will exploit 
this knowledge in our favour to decouple the scalar field 
fluctuation $\varphi$ from the system. 

Under this decoupling approximation $m_1^2 \gg \Box$, 
the equation of motion for field $\varphi$ acquires a simplified expression.  
Keeping only the leading order terms we get
\beq
\label{eq:largeM1_1}
\varphi = -\frac{1}{3}\kappa - \frac{1}{m_1^2} \left(
\Box \chi - \frac{(\xi_{12} + \xi_{21})R \chi}{2} 
\right) \, . 
\eeq
The other equation of motion for the field $\chi$ gets similarly 
simplified when the reduced equation of motion for $\varphi$ is 
plugged in it. This is given by,
\beq
\label{eq:largeM1_2}
\left(-\Box + \xi_2 R - \frac{4}{9} m_2^2 \right) \chi 
+ \frac{(\xi_{12} + \xi_{21})R}{3} \kp = 0 \, ,
\eeq
where higher order non-linear terms are ignored. This residual linear 
equation can be solved easily by inverting the operator, 
\beq
\label{eq:resiEQM}
\chi = -\frac{\kp (\xi_{12} + \xi_{21})}{3} 
\left(-\Box + \xi_2 R - \frac{4}{9} m_2^2 \right)^{-1} R \, . 
\eeq
This appears as a constraint in the system after decoupling of the 
scalar mode $\varphi$ has occurred. Here the inversion acts on the Ricci scalar $R$. 
One can then plug the solution for $\varphi$ from Eq. (\ref{eq:largeM1_1})
back in the action of the theory, given in Eq. (\ref{eq:symbreakAct}). 
This generates leading and sub-leading terms in the action. Under the 
decoupling approximation one can safely ignore the sub-leading part 
(which are of {\cal O}($1/m_1^2$)), as the dominant role will be played 
by leading part of the action. The leading part of the action is given by,
\bea
\label{eq:residual}
S &=& \int \, {\rm d}^4x 
\sqrt{-g} \biggl[
- \frac{4}{81} m_1^2 \kp^2 + \frac{2 \xi_1}{9} \kp^2 R 
+ \frac{\xi_{12} + \xi_{21}}{3} \kp R \chi 
\notag \\
&&
+ \frac{1}{2} (\pt \chi)^2 + \frac{\xi_2}{2} R \chi^2
- \frac{2}{9} m_2^2 \chi^2
-\frac{1}{4} \lam_2^2 \chi^4 
\biggr]\, .
\eea
This is the action where the decoupling of field $\varphi$ has occurred.
This residual action contains only terms which are dominant after the 
decoupling (ignoring the contribution from {\cal O}($1/m_1^2$)).
Here the field $\chi$ is coupled non-minimally 
with the background space-time. In the process of symmetry breaking 
a large cosmological constant is generated which remains present in the 
residual action for the field $\chi$. However, the effects of this large cosmological 
constant gets shielded if the gravitational coupling dictating the behaviour of
metric under the influence of cosmological constant, goes to zero. This 
situation actually occurs in the current case of induced gravity coupled with 
higher-derivatives (see Eq. (73), Fig. 7, Fig. 11 and Fig. 13 of \cite{Narain2016}). 
Here it is noticed that renormalization group running of induced Newton's constant 
is such that it goes to zero in deep infrared. This behaviour of 
gravitational Newton's constant was also witnessed in pure 
higher-derivative gravity \cite{NarainA1,NarainA2} (and in higher-derivative 
gravity coupled with gauge fields \cite{Narain:2012te,Narain:2013eea}). 
In then implies that in infrared as the gravitational coupling strength weakens, the 
back-reaction of vacuum energy gets severely shielded making the large 
cosmological constant innocuous. In the following we will therefore 
ignore the generated cosmological constant term.

The field $\chi$ slow rolls for the case when 
coupling $\lam_2$ and mass $m_2^2$ is small. In which case the kinetic 
term of the field $\chi$ can be ignored. This action has a simple form given by,
\bea
\label{eq:residual}
S &=& \int \, {\rm d}^4x 
\sqrt{-g} \biggl[
\frac{2 \xi_1}{9} \kp^2 R + \frac{\xi_{12} + \xi_{21}}{3} \kp R \chi 
+ \frac{\xi_2}{2} R \chi^2
\notag \\
&&
- \frac{2}{9} m_2^2 \chi^2
-\frac{1}{4} \lam_2^2 \chi^4 
\biggr]\, .
\eea
In this slow-roll regime the field $\chi$ no longer has dynamics. It couples 
with the background curvature driving the dynamics of space-time. In this 
if the following approximation holds
\beq
\label{eq:finetune}
\frac{2\xi_1}{3(\xi_{12} + \xi_{21})} \gg 
\frac{\chi}{\kp} \gg \frac{2(\xi_{12} + \xi_{21})}{3\xi_2} \, 
\eeq
then the linear $\chi$ term doesn't contribute compared to quadratic 
and quartic pieces and the induced Einstein-Hilbert piece. 

The approximation in Eq. (\ref{eq:finetune}) is not unreasonable 
in the case of induced gravity coupled with higher-derivative gravity \cite{Narain2016}. 
This model is a perturbatively renormalizable model in four dimensional space-time 
where the RG analysis does not put any 
constraint on the required values of the 
parameters $\xi_1$, $\xi_2$, $\xi_{12}$ and $\xi_{21}$. This means 
that these parameters can be chosen freely. Exploitation of 
this freedom into choosing those values of the parameters 
$\xi_1$, $\xi_2$, $\xi_{12}$ and $\xi_{21}$ such that the condition 
stated in Eq. (\ref{eq:finetune}) is fulfilled, suppresses the linear term 
in $\chi$ in the residual action stated Eq. (\ref{eq:residual}). 
This will then leave us with the 
induced Einstein-Hilbert term, and quadratic terms in the 
fields plus the interaction pieces. 
This action is
\bea
\label{eq:residualact1}
&&
S = \int {\rm d}^4x \sqrt{-g} \biggl[
\frac{2 \xi_1}{9} \kp^2 R + \frac{\xi_2}{2} R \chi^2
- \frac{2}{9} m_2^2 \chi^2
\notag \\
&&
-\frac{1}{4} \lam_2^2 \chi^4 
\biggr] \, .
\eea
This action of the field $\chi$ is achieved once the decoupling of the 
heavy scalar has occurred, and the field $\chi$ starts to slow roll
where it satisfies the condition Eq. (\ref{eq:finetune}). This results in 
a non-dynamical action of $\chi$ field which couples with the background 
dynamical geometry. This action comes along with the constraint imposed in the 
Eq. (\ref{eq:resiEQM}). This constraint when plugged back into 
the residual action leads to the non-local action of the theory
\bea
\label{eq:SNLred}
&&
S_{\rm NL} =  \int {\rm d}^4x \sqrt{-g} \biggl[
\frac{2 \xi_1}{9} \kp^2 R 
\notag \\
&&
- \frac{2 m_2^2 \kp^2 (\xi_{12} + \xi_{21})^2}{81} 
R \left(-\Box + \xi_2 R - \frac{4}{9}m_2^2\right)^{-2} R 
\notag \\
&&
+ \frac{\xi_2 \kp^2 (\xi_{12} + \xi_{21})^2}{18} 
\notag \\
&&
\times
R \left(-\Box + \xi_2 R - \frac{4}{9}m_2^2\right)^{-1} 
R \left(-\Box + \xi_2 R - \frac{4}{9}m_2^2\right)^{-1} R
\notag \\
&&
- \frac{\lam_2^2 \kp^4}{324} (\xi_{12} + \xi_{21})^4 \bigl\{
\left(-\Box + \xi_2 R - \frac{4}{9}m_2^2\right)^{-1} R
\bigr\}^4
\biggr] \, . 
\eea
Here the first term is the induced gravitational term 
while other terms are generated under decoupling and 
approximations. This can be seen as a heuristic derivation of the 
non-local action that has been studied extensively in 
\cite{Cusin2016nzi,Maggiore2016gpx,Cusin:2015rex,Dirian:2016puz}
and has been argued to reproduce dark energy data 
as good as $\Lam$-CDM. 

If we define the induced Newton's constant as 
$9/(32\pi G)^{-1} = \xi_1 \kp^2$, then one can pull 
out the factor of Planck's mass and write the action 
in a form which can be compared to the existing models 
given in \cite{Cusin2016nzi,Maggiore2016gpx,Cusin:2015rex,Dirian:2016puz}. 
This will become 
\bea
\label{eq:NLactMM}
&&
S_{\rm NL} =  \frac{M_p^2}{2} \int {\rm d}^4x \sqrt{-g} \biggl[
R 
\notag \\
&&
- \frac{\mu^2}{6} R \left(-\Box + \xi_2 R - \frac{4}{9}m_2^2\right)^{-2} R 
\notag \\
&&
+ \rho_1^2 R \left(-\Box + \xi_2 R - \frac{4}{9}m_2^2\right)^{-1} 
\notag \\
&&
\times
R \left(-\Box + \xi_2 R - \frac{4}{9}m_2^2\right)^{-1} R
\notag \\
&&
- \rho_2^2 \left\{\left(-\Box + \xi_2 R - \frac{4}{9}m_2^2\right)^{-1} R\right\}^4
\biggr] \, ,
\eea
where $M_P^2 = (4\xi_1 \kp^2)/9$ is the reduced Planck's mass and
\bea
\label{rho1rho2}
&& \mu^2 = \frac{2 (\xi_{12} + \xi_{21})^2 m_2^2}{3\xi_1} \, , \notag \\
&& \rho_1^2 = \frac{\xi_2 (\xi_{12} + \xi_{21})^2}{4 \xi_1} \, , 
\hspace{3mm}
\rho_2^2 = \frac{\lam_2^2 M_P^2 (\xi_{12} + \xi_{21})^4}{32 \xi_1^2} \, ,
\eea
where these entities are defined to make connection with the 
non-local model defined in Eq. (\ref{eq:nonLocalconfGR}). 
It was observed in \cite{Cusin2016nzi,Maggiore2016gpx,Cusin:2015rex,Dirian:2016puz}
that the parameter $\mu$ should be a very small quantity 
(of the ${\cal O}(H_0)$), in order to explain current accelerated 
expansion of the Universe. Following their work, it is seen 
one can still expect reasonable values of $m_2^2$ 
by appropriately choosing $\xi_1$ and $(\xi_{12} + \xi_{21})$. 
The freedom allowed by the renormalizability of the theory allows one to choose these 
factors freely. 

At this point one can make an order of estimate for the parameter
$m_2^2$ by using the values of $\mu^2$ considered in 
\cite{Cusin2016nzi,Maggiore2016gpx,Cusin:2015rex,Dirian:2016puz}.
From their analysis it is known that $\mu$ is of the order of current Hubble radius 
$H_0$. This is a very small quantity ${\cal O}(10^{-32})$ eV.
The authors of \cite{Cusin2016nzi,Maggiore2016gpx,Cusin:2015rex,Dirian:2016puz}
further mentions that even though $\mu$ (in their model) is a small quantity, but 
it doesn't describe the fundamental length scale that enters the 
physical system. This fundamental length scale is $\Lam_{RR}$
and is related to $\mu^2$ via $M_P^2$ in the following manner,
\beq
\label{eq:LamRR}
\Lam_{RR}^4 = \frac{M_P^2}{12} \mu^2 \, .
\eeq
For our present case this will imply that $\Lam_{RR}$ is related to 
the $m_2$ and $\kp$ in the following way
\beq
\label{eq:LamRR1}
\Lam_{RR}^4 = 
\frac{2}{81} (\xi_{12} + \xi_{21})^2 m_2^2 \kp^2 \, .
\eeq
According to the estimates made in works 
\cite{Cusin2016nzi,Maggiore2016gpx,Cusin:2015rex,Dirian:2016puz}, it is seen 
that for the reduced Plank mass $M_P$, if $\mu \sim {\cal O}(H_0)$ then 
$\Lam_{RR} \sim 10^{-3}$ eV. In GUT (Grand Unified Theory models) scenarios,
if the scale symmetry broke around GUT scale resulting in generation of 
induced Newton's constant, then this will imply that 
$\kp \sim 10^{16}$ GeV and correspondingly $\xi_1 \sim 100$. 
This knowledge further leads to the mass of the scalar field $\chi$ ($m_2^2$)
following Eq. (\ref{eq:LamRR1}), which gives 
$m_2 \sim 10^{-30} (\xi_{12} + \xi_{21})^{-1}$ eV. 
Exploiting the freedom offered in the choice of values of 
parameters $(\xi_{12} + \xi_{21})$, allows one
to have a reasonable $m_2$. This particular model which has a well-defined 
UV completion and is free of ghosts can be seen to offer an 
interesting picture where low-energy non-local interaction 
emerges leading to accelerated expansion in late 
time Universe.

\section{Conclusion}
\label{conc}

We tried to understand dark energy and the accelerated 
expansion it causes at late times of the Universe. We present a local unified model 
of a modified theory of gravity with a coupled system of scalar fields in Eq. (\ref{eq:coupleAct}). 
This scale-invariant model has a well-defined UV behaviour in the 
sense that the theory is perturbatively renormalizable to 
all loops \cite{Stelle77, Fradkin1981, Fradkin1982} and 
in tachyons free regime higher-derivative ghosts have been excised out 
\cite{Salam1978,Julve1978, NarainA1, NarainA2, Narain2016}. In this 
model we would like to seek whether it can explain late time cosmic acceleration 
observed in the Universe. We considered a two coupled real scalar field model
which also interacts non-minimally with gravity. One of the scalar acquires a
VEV and in turn generates Newton's constant and masses for the fields. 
This scalar gets decoupled from the system as its mass becomes 
very large in the infrared, leaving behind a simple system. 

Although breaking of scale-symmetry generates cosmological constant in the theory
but its effect are shielded from affecting the dynamics of space-time due to the 
weakening of induced gravitational coupling constant in the infrared. 
This weakening of induced gravitational coupling is indeed observed in 
\cite{Narain2016} (and for Einstein-Hilbert coupled with higher-derivates 
in \cite{NarainA1, NarainA2, Narain:2012te,Narain:2013eea}), where such infrared
vanishing of Newton's constant shield the effect of large cosmological constant. 
For the case of small coupling $\lam_2$ and small mass, 
the field $\chi$ starts to slow roll, implying that its kinetic term can be ignored
in the action. At this point the action for field $\chi$ consist of only non-dynamical 
pieces and their interaction with the space-time. The RG running of the 
of parameters as seen in \cite{Narain2016} is such that the approximation 
stated in Eq. (\ref{eq:finetune}) is satisfied if one chooses large values of 
$\xi_1$ and $\xi_2$. This doesn't hampers the renormalizability as explained in 
\cite{Narain2016}. Under this condition the linear term in $\chi$ can be ignored 
over the quadratic and quartic piece. 
The integrating out of this leftover 
scalar from the simplified action leads to a non-local 
version of theory whose leading term matches the 
non-local gravity action studied extensively by 
\cite{Cusin2016nzi, Maggiore2016gpx, Barreira:2014kra, Dirian:2014bma, Dirian:2016puz}
where it was noticed that it explain the late-time cosmic acceleration as well 
as $\Lam$-CDM. This local scale-invariant system is then 
a good model which not only has a good UV completion 
but also leads to non-local gravity theory which can explain 
dark energy. In the papers \cite{Cusin2016nzi, Maggiore2016gpx, 
Barreira:2014kra, Dirian:2014bma, Dirian:2016puz} the authors 
introduced a controllable parameter $\mu^2$ (or $\Lam_{RR}$)
whose value was fixed by matching with the dark-energy data. However, 
an explanation for a possible origin of this length-scale was lacking. 
Here we derive the parameter 
$\mu^2$ (or $\Lam_{RR}$) in terms of parameters present 
in the local model which is given in Eq. (\ref{eq:LamRR1}). This
indicates a possible origin for length scale $\mu^2$ (or $\Lam_{RR}$). 

Generalised non-local gravity actions are also favourable 
in the sense that they offer super-renormalizability of the full theory, implying 
that the quantum theory is free of UV divergences 
\cite{Moffat:2010bh,Biswas:2010zk,Modesto:2011kw, Briscese:2013lna, Modesto:2014lga}.
The non-locality present in these theories provides an extra suppression factor 
in the propagator at high-energies leading to a well defined 
UV behaviour. Such non-local theories also arise at low energy from 
a more fundamental UV complete theory such as string theory. It will be 
interesting to see whether the model proposed in this paper 
can be elegantly extended to include these generalised 
non-local theories. This will be considered in a future 
publication. 

The model presented here is UV complete (perturbatively renormalizable to all loops
\cite{Stelle77, Fradkin1981, Fradkin1982} where ghosts are evaded 
\cite{Salam1978,Julve1978, NarainA1, NarainA2, Narain2016}) and approximations
have been made in an attempt to arrive at the non-local form of the action. 
Perhaps an inclusion of additional symmetry incorporated 
at the level of local action might be able to give better handle 
over the scalar sector of the theory where the assumptions 
considered here will naturally arise. It may be 
possible that the residual scalar $\chi$ is a composite of 
some fundamental fermions, in which case dark energy arises 
when the condensate is formed \cite{Bhatt:2009wb}. This scenario 
will be explored in the future. Moreover, it will be interesting to 
investigate the modification of the behaviour of infrared gravitational 
field theory (classical and quantum) under the inclusion of such non-localities. 
In particular the effect on gravitational waves (GW) created by such infrared 
non-local modification of gravity is worthy of investigation. Using 
future GW detectors one can possibly test such models
\cite{Belgacem:2017kev,Belgacem:2017ihm}. This will be 
presented in future publication.

\bigskip
\centerline{\bf Acknowledgements} 

We would like to thank Nirmalya Kajuri and Nick Houston for useful discussions 
during the course of this work. This research was supported  by the 
Projects 11475238 and 11647601 supported by the National Natural Science 
Foundation of China, and by the Key Research Program of Frontier Science, CAS.



\end{document}